# Magnetophotonics for sensing and magnetometry towards industrial applications

Conrad Rizal[1], Maria Grazia Manera[2], Daria O. Ignatyeva[3,4,5], J. R. Mejía-Salazar[6], Roberto Rella[2], Vladimir I. Belotelov[3,4,5], Francesco Pineider[7], and Nicolò Maccaferri[8,9]

[1]Seed NanoTech International Inc., 4 Howell St., Brampton, ON, L6Y 3J6 Canada

[2]IMM-CNR Institute for Microelectronics and Microsystems, strada prov.le Lecce-Monteroni, c/o campus Ecotekne, 73100 Lecce, Italy

[3]Russian Quantum Centre, Skolkovo IC, Bolshoy Bulvar 30, bld. 1, 143025 Moscow, Russia;

[4]Vernadsky Crimean Federal University, pr-kt Vernadskogo 4, 295007 Simferopol, Russia;

[5]Photonic and Quantum Technologies School, Lomonosov Moscow State University, Leninskie gori, 119991

[6]National Institute of Telecommunications (Inatel), 37540-000, Santa Rita do Sapucaí, MG, Brazil

[7]Department of Chemistry and Industrial Chemistry, University of Pisa, Via G. Moruzzi 13, 56124 Pisa, Italy

[8]Department of Physics and Materials Science, University of Luxembourg, 162a Avenue de la Faïencerie L-1511 Luxembourg, Luxembourg

[9]Department of Physics, Umeå University, Linnaeus väg 20, SE-90736 Umeå, Sweden

Magnetic nanostructures sustaining different kinds of optical modes have been used for magnetometry and label-free ultrasensitive refractive index probing, where the main challenge is the realization of compact devices able to transfer this technology from research laboratories to smart industry. This Perspective discusses the state-of-the-art and emerging trends in realizing innovative sensors containing new architectures and materials exploiting the unique ability to actively manipulate their optical properties using an externally applied magnetic field. In addition to the well-established use of propagating and localized plasmonic fields, in the so-called magnetoplasmonics, we identified a new potential of the all-dielectric platforms for sensing to overcome losses inherent to metallic components. In describing recent advances, emphasis is placed on several feasible industrial applications, trying to give our vision on the future of this promising field of research merging optics, magnetism, and nanotechnology.

**I. Introduction**

Beyond its fundamental scientific importance, manipulating light at the nanoscale, i.e. nanophotonics, can enable many real-life applications, including energy harvesting and photovoltaics, wave-guiding and lasing, optoelectronics, biochemistry and medicine. To achieve new functionalities, the combination of nanoscale electromagnetic (EM) fields with other materials properties has become increasingly appealing. In this framework, magnetophotonics is an emerging area aiming to combine magnetism, magneto-optical (MO) effects, and nanophotonics to find new efficient ways of controlling light-driven phenomena using magnetism and vice-versa [1-9]. Important areas of application of magnetophotonic structures include bio- and chemical sensing/imaging [10], and heating/theranostic [11, 12]. The realization of optimized transducers able to monitor very small (up to $10^{-6}$) refractive index (RI) changes at the interface with investigated media is still a challenge in the sensing field. Detection of very low (up to the attomole) concentrations of small molecules is a challenge in different biological, medical, industrial contexts, and a few directions of possible improvements were attempted. The potential offered by magnetophotonics to manipulate light below the diffraction limit by an external agent, i.e., a magnetic field, has been proved to allow fine engineering of the distribution of local field on the transducers surfaces that, together with strong confinement of EM energy over small (few $nm^3$) volumes, ensure high sensing abilities at the single-molecule level. New transducing platforms with amplified sensing signal probes and interrogation schemes have been proposed in the past years, demonstrating a much better resolution in following all sensing (e.g. molecular adsorption on a surface of a molecule or a virus) events at the transducer interfaces. Many works have appeared during the past two decades on this topic, showcasing the potential advantages of magnetic modulation to increase the performance of nanophotonic sensing devices. Generally, the active and reproducible control of an optical resonance can afford a dramatic boost in signal tracking through synchronous detection. Miniaturization and integration with mechanical hardware as well as with microfluidic tools for the realization of smart and point of care devices will be the real challenge to be overcome. In this sense, a game changer can be traced in the nanophotonic-based sensing field.

In this Perspective, we try to provide the most extended overview on this rising research area and give our vision on the future of this field of applied physics. We divide the Perspective into seven sections. Section I contains a brief introduction to the Perspective and how it has been organized and conceived by the authors. Section II mainly covers magnetophotonic nanostructures made of metallic materials (noble metals and magnetic) supporting plasmonic excitations, that is, collective oscillations of conduction electrons resonantly excited by electromagnetic radiation. Part (a) of Sec. II presents the overview of fundamental works in magnetoplasmonic sensing by using surface propagating plasmons (SPPs). In Sec. II (b) we introduce the reader to nanostructures supporting localized surface plasmons (LSP), which have been used to detect both volumetric refractive index changes and molecular adsorption processes. In both Secs. II (a) and II (b) we discuss the fundamental limitations and the main strategies used to maximize the MO sensing capabilities of the devices. In Sec. III, we discuss all-dielectric nanostructures supporting Tamm surface modes, and how we can use them to improve the performances of MO sensors for the detection of many biological entities, for instance, inside a cell. Particular emphasis is put in the description of the physics of these structures since they are quite uncommon, and only recently were they proposed as promising, novel platforms to advance the magnetophotonic sensing research field. Section IV showcases new materials and hybrid systems used in magnetophotonic sensing to achieve higher detection capabilities. In Sec. V we provide an overview of the potential use of magnetophotonic devices for magnetometry purposes. Sec. VI brings an overview of the industrial landscape where the devices introduced in this Perspective can have a significant impact. We conclude by giving our outlook on the field and summarizing the recent advances that pave the way to practical magnetophotonic devices for sensing.

**II. Magnetoplasmonic nanostructures**

Nanoscale materials supporting the excitation of surface plasmon (SP) modes upon interaction with light have attracted attention due to their unique properties. Strong localization and enhancement of electric field are achieved at their surfaces within subwavelength volumes, which can be tailored by proper nanostructure design, arrangement and dielectric environment of the nanostructures surfaces. Besides enabling amplification of different optical phenomena, such as surface spectroscopies or MO phenomena, plasmonic properties allow probing minute changes in the dielectric properties of the media in which surface plasmon waves are excited. Both SPP and LSP resonances (see **Figure 1**) have been investigated in this context by exploiting their peculiar optical properties mainly bound to the strength and intensity of the field localization and the extent of plasmon decay lengths. The advantage is the possibility of monitoring in real-time any kind of local change in refractive index (RI) occurring within the decaying plasmonic field without the use of an additional tag. In a typical sensing experiment, this can be due to surface adsorption of (bio)molecules or to volumetric changes in gas or liquid media close to the metal surface. A proper combination of metal nanostructure design, efficient nanofabrication methods, surface functionalization protocols and accurate detection schemes ensures the achievement of essential parameters for label-free sensing platforms, namely sensitivity, specificity, low cost, miniaturization capabilities and integration with microfluidic tools. Several applications have been proposed, such as detection of DNA sequences [13], sensing of pathogens [14], real-time interaction between biomolecules and drug compounds [15]. Applications in detecting gas or volatile organic compounds (VOCs) have been proposed for environmental monitoring [16,17].

In its most basic form, the sensor configuration can include prism materials, various types of substrates, optimized sensor configuration (*Kretschmann* or *Otto* [18]), and variable thicknesses of the analyte (e.g., gas, semi-liquid, or liquid media). The signal is retrieved as a function of the wavelength or the incidence angle of the light source, as the magnetic field is tuned [19]. Transverse, polar, and longitudinal MO configurations can be employed to detect chiral molecules, for example, by studying the rotation and the ellipticity of the transmitted light [20,21]. Research in this field is oriented toward determining proper plasmonic modes able to respond to the needs of the investigated sensing problem. In fact, important results have been achieved, thanks to the development of the fields as mentioned above. Further insight in near and far-field coupling has been obtained, with significant benefit in sensing performance. However, huge efforts are still needed to improve the sensing performance in plasmonic-based transducers. The challenge is detecting minute RI changes due to small molecules' surface binding events or to the dispersion of small molecules in low concentration solutions. The necessity of extreme sensitivity, even down to the single-molecule level, is due to the interest in various biological, chemical, and medical investigations [22-,24]. The goal is the development of new sensing strategies that maintain all the advantages of conventional plasmonic sensing and, at the same time, can add smart functionalities for reliable point of care devices.

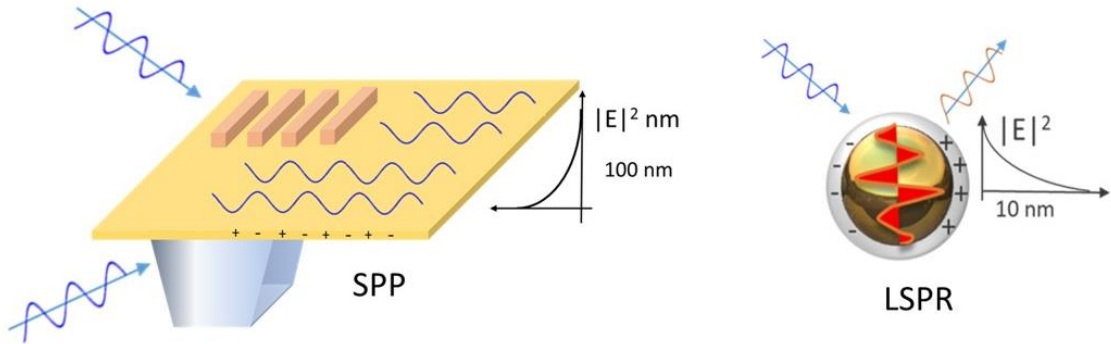

**Figure 1.** Left panel: typical techniques used to excite SPP waves on metallic thin films. The propagation length of these waves is few microns, and the extension along the z-direction is few hundreds of nm. Right side: a sketch of a localized surface plasmon resonance in a metallic nanostructure. In this case, the extension of the evanescent field is of the order of few tens of nm. Beyond this distance, the nanostructure behaves like an antenna and can couple to other structures.

Before reviewing this research field, it is necessary to define the analytical parameters characterizing sensing performances of plasmonic transducers to compare them in different platforms. Sensor response is the main parameter to be considered: depending on the interrogation scheme, spectral, angular, intensity, or polarization, changes of transmittance /reflectance spectra can be selected for monitoring changes of plasmon resonance condition. Intensity-interrogation methods are attractive as they allow a greater integration in high throughput platforms such as SPP imaging and microfluidics tools. Despite possessing lower RI resolution compared to other detection schemes, they ensure a faster response to RI changes, representing an essential analytical tool for real-time detection. The simplest quantitative performance parameter of a plasmonic sensor is the sensitivity to bulk refractive index changes (bulk sensitivity, $S_B$). It is defined as the slope of the sensor response in the investigated RI range. Surface sensitivity ($S_s$), intended as the sensor response to a determined number of molecules adsorbed to the surface, quantifies the ability to monitor molecular adsorption events of particular interest for biosensing purposes [25]. In order to take into account also the spectral shape of the plasmonic response, some authors also introduced a Figure of Merit (FoM) as an additional analytical parameter. Defined as the ratio of Sensitivity (either $S_B$ or $S_s$) to the full-width at half maximum (FWHM) of the curve, it is of particular interest when comparing different plasmonic platforms where spectral or angular shift are chosen as a sensor response. Improvements in FoM can be achieved by enhancing sensitivity and reducing the FWHM of the LSP spectral curve. The first goal can be reached by proper nanostructure design in terms of material's choice, size, shapes, and arrangements. The FWHM of the spectral resonance curve can be reduced by lowering inherent losses in metallic nanostructures, thus reinforcing the strength of the resonance. Further attempts to accomplish this goal can be found in the literature: stronger EM enhancements due to near field resonant coupling as well as the interaction between plasmonic and optical modes are among the most prevailing.

FoM is not the only analytical parameter for comparing sensing performances of plasmonic transducer platforms, as it does not consider instrument factors such as signal noise that depends on the detection method and overall instrumentation performance. Sensor resolution (RES) and limit of detection (LOD) are then the most appropriate parameters for accomplishing this role. Resolution is obtained from the noise in the detector output $\sigma_{so}$, and $S_B$ according to RES = $S_B/\sigma_{so}$. LOD has a connotation familiar to analytical chemists, and it is defined as the minimum variation of the measured quantity that the sensor can detect with reasonable certainty (generally intended as 3 or 5 times the standard deviation of the signal). Improving RES can then be achieved by reducing noise and enhancing the sensitivity of the sensing platform. The above parameters allow the determination of the smallest detectable change in RI or minimum detectable coverages if bulk or surface sensitivity are investigated, respectively. Phase-sensitive detection schemes and active modulation techniques are other possible strategies for improving the signal-to-noise ratio (SNR) and enhancing LOD. Mechanical or phase-modulated SPP sensors are some examples of experimental configurations exploiting external modulation techniques [26, 27]. The design and implementation of configurations where an external agent, such as temperature, electric or magnetic fields, can control surface plasmons is known as Active Plasmonics, and constitutes the basis for innovative optoelectronic components. Among different routes proposed so far, the use of a magnetic field seems a very promising one.

**IIa. Magnetoplasmonic sensing with SPPs**

The so-called magneto-plasmonic (MP) modulation arises from the simultaneous excitation of MO effects and the SPP in structures with plasmonic and MO activity. Light localization associated with excitation of SPPs can be used for resonant enhancement of MO signals in properly arranged nanostructures. In this mechanism, the enhanced EM field associated with SPPs excitation in plasmonic nanomaterials is also distributed inside an adjacent magnetic layer, thus increasing its MO activity. Excitation of propagating SPP modes can be used for resonant enhancement of the different MO effects both in Reflectance (MOKE) or in Transmission (Faraday) either in Transverse, Polar, or Longitudinal experimental configurations [28]. In the following, we will offer examples from recent literature where the magnetoplasmonic approach is exploited in different combinations with surface plasmon mode excitations to overcome and surpass all limitations of traditional plasmonic sensing, also allowing for developing smart and integrated devices. The sensing process of the MO sensor can be realized both via the spectral measurement of the shift of the transverse MO Kerr effect (TMOKE) resonance position or via the measurement of the TMOKE value variation for the fixed near-resonant angle and wavelength.

The first investigation towards a MO-SPP-based biosensor with enhanced sensitivity is compared to standard SPP-based sensors used in combining TMOKE and SPP effect in *Kretschmann* configuration [29]. Multilayers composed of ferromagnetic (FM) and noble metals were proposed as transducing elements as they combine the large MO activity of FM materials and the exceptional plasmonic properties of noble metals. A remarkable MO signal was recorded, based on the non-reciprocal variation of the SPP wave vector with an external magnetic field placed perpendicular to the plane of incidence and parallel to the multilayer surface. In addition to the excitation conditions of the SPPs, the MO enhancement depends on the refractive index of the dielectric in contact with the metal layer; therefore, the plasmons excited by the incident light and modulated by the external magnetic field give rise to a new modulated sensing probe. By exploiting the modulated nature of the signal and the sharp Fano-like resonant plasmonic enhanced MO signal, resolving small refractive index (RI) changes at the metal-dielectric interfaces is easily achievable. A threefold improvement in bulk sensitivity with respect to traditional plasmonic based sensing devices could be demonstrated.

Since the above-mentioned first paper, different research groups have worked in defining the optimal configuration for sensing and biosensing [30]. Enhancement of MO properties results from a fine balancing between optical absorption and MO activity, as well as from thickness and position of metal and FM layer in the sandwich nanostructure. Upon a proper choice of metal and FM materials, the resulting TMOKE signal, described as a relative intensity change of the reflected light, is strongly amplified when the SPP excitation condition is satisfied. Different metallic layer combinations have been investigated both experimentally and theoretically from a structural, optical and MO point of view (Co/Au, Ag/Co/Ag, Au/Co/Au, and Au/Fe/Au [31,32]). Some of them have also been tested as MP transducers for RI changes or molecular adsorption events monitoring. Application of this new transducing concept barely as RI transducers is reported in the literature [33-36]. The possibility of coupling the novel MP sensing ability with the chemical and biological functionalization of the transducers has also been demonstrated in a novel class of chemical gas [37-42] and biological sensing devices [43]. These investigations have shown that the thickness of the FM material is critical due to their high absorbance, so that thickness in the multilayers system is important.

A first attempt to overcome the problem was reported by Kubler et al. [44] where an exchange biased (EB) layer system, a combination of an FM and an anti-FM (AFM) layer with a remnant magnetization of the FM layer. The MO activity of the optimized EB systems, with a proper Au capping layer, is demonstrated comparable to the Au/Co/Au transducing trilayer for the above-mentioned MO-SPP-based sensors. Even if still explored, this class of materials could offer further fascinating insight due to the possibility of being employed for the transport of superparamagnetic particles as biomolecular carriers [45]. Several approaches have been proposed for enhancing sensitivity performances of MP multilayers supporting propagating SPPs. Ignatyeva et al. proposed a new MO heterostructure composed of a thin cobalt layer, a gold layer and a specially designed photonic crystal (PC) supporting ultralong surface plasmon propagation [46]. The MO response is guaranteed by the cobalt layer, while the quality factor of the plasmon resonance is determined by the PC and the proper choice of thicknesses of the gold and cobalt layers. An unprecedentedly narrow angular TMOKE resonance width is thus achieved, allowing to record detection limits in terms of RI units more than one order of magnitude higher than previously reported.

Finding miniaturizable and integrable magnetoplasmonic platforms for biosensing still remains one of the major challenges in SPP biosensing. The use of grating couplers for SPP excitation can face up the issue very well. Avoiding prism-coupling for plasmon modes excitation offers the advantage of easier miniaturization and integration in nanoscale devices. Proper combination with microfluidics systems for real-time analysis of biomolecular kinetics make them a smart choice for the realization of portable point of care devices. Moreover, by exploiting the extraordinary optical transmission originating from the resonant excitation of SPPs in periodically patterned nanostructures, it has been shown superior sensing performances and different advantages over the traditional prism-coupled plasmonic sensors [47,48].

FM and noble metal nanolayers have been combined in one-dimensional (1D) periodic magnetoplasmonic gratings to realize an ultrasensitive MO-SPP device: Fedyanin et al. designed a 1D MP grating where the combination of Ag/Fe layers is proposed and TMOKE signal investigated as MO sensing signal. Optimization of the FM layer thickness resulted in a minimum detectable signal of $5.6 \times 10^{-4}$ RIU [49]. Even better results are reported by Lee et al: here the grating structure, made of a hybridized layer Au/Fe/Au multilayer, is still investigated by TMOKE signal resulting in a minimum detectable value of $7.6 \times 10^{-6}$ RIU with the possibility also to monitor biomolecular surface interactions with limit of detection in the order of the nanomolar [50]. A novel 1D magnetoplasmonic architecture is proposed theoretically by Diaz-Valencia et al. [51]. By a proper choice of materials and their geometrical arrangement, i.e. an Au grating grown on a MO metallic substrate, excitation of surface plasmon resonances is mainly localized at the sensing layer, where interaction with the analyte occurs. Enhanced TMOKE signals with very narrow Fano-like resonant peaks are thus achieved, demonstrating an extreme sensitivity to refractive index changes of the surrounding media with higher resolution with respect to reflectance curves. Two-dimensional (2D) metal nanohole arrays have also attracted the interest of the scientific sensing community. Making use of the extraordinary optical transmission phenomenon which originates from the resonant excitation of SPPs in these periodically patterned nanostructures, Blanchard-Dionne and co-workers obtained sensing performances and exploiting different advantages over the traditional prism-coupled plasmonic sensors [52]. Their MO sensing performances were proposed in a recent theoretical paper by Caballero et al., where an array of metallic nanoholes in a trilayer of Au/Co/Au is investigated in a TMOKE configuration [53]. Their theoretical work reported bulk figures of merit that are two orders of magnitude larger than those of any other type of plasmonic sensor. The results are promising for MP sensing devices. Recently, Li and colleagues theoretically demonstrated an MP sensor composed of Au/Co bilayer nanodisk arrays on top of optically thick metallic films, which supports a narrow SPP with a bandwidth of 7 nm and allows for refractive index sensitivities as high as 717 nm/RIU [54].

Pure FM nanogratings have been also used to tune the plasmonic resonance strength, and consequently its potential sensing capabilities, by modulating the nanograting cross-section [55]. Chichelero and co-workers also showed that, by a narrow, asymmetrical Fano-shape resonance, TMOKE configuration results are more useful for sensing purposes, promising excellent resolution of small refractive index changes of the dielectric media [56]. Interesting potentialities for sensing applications can be found also in the hole-array metamaterial platforms made of $Ni_{81}Fe_{19}$/Au multilayers proposed by Armelles et al. [57]. The mid-IR plasmon response associated to the proposed square lattice array experiences large intensity modulation under the application of a very small magnetic field. This can open the route to exciting perspectives for MP sensing in unexplored wavelength ranges. The optimal combination of magnetic properties, namely: low magnetic anisotropy and smaller magnetic damping among FM metals, make the choice of permalloy materials promising not only as chemical sensing transducing platforms but also in the field of physical sensors. Belyaev et al. obtained an MP crystal with a permalloy layer as a magnetic sensor whose sensing performances correlated with magnetic and MO properties of MP crystals through their TMOKE signal [58].

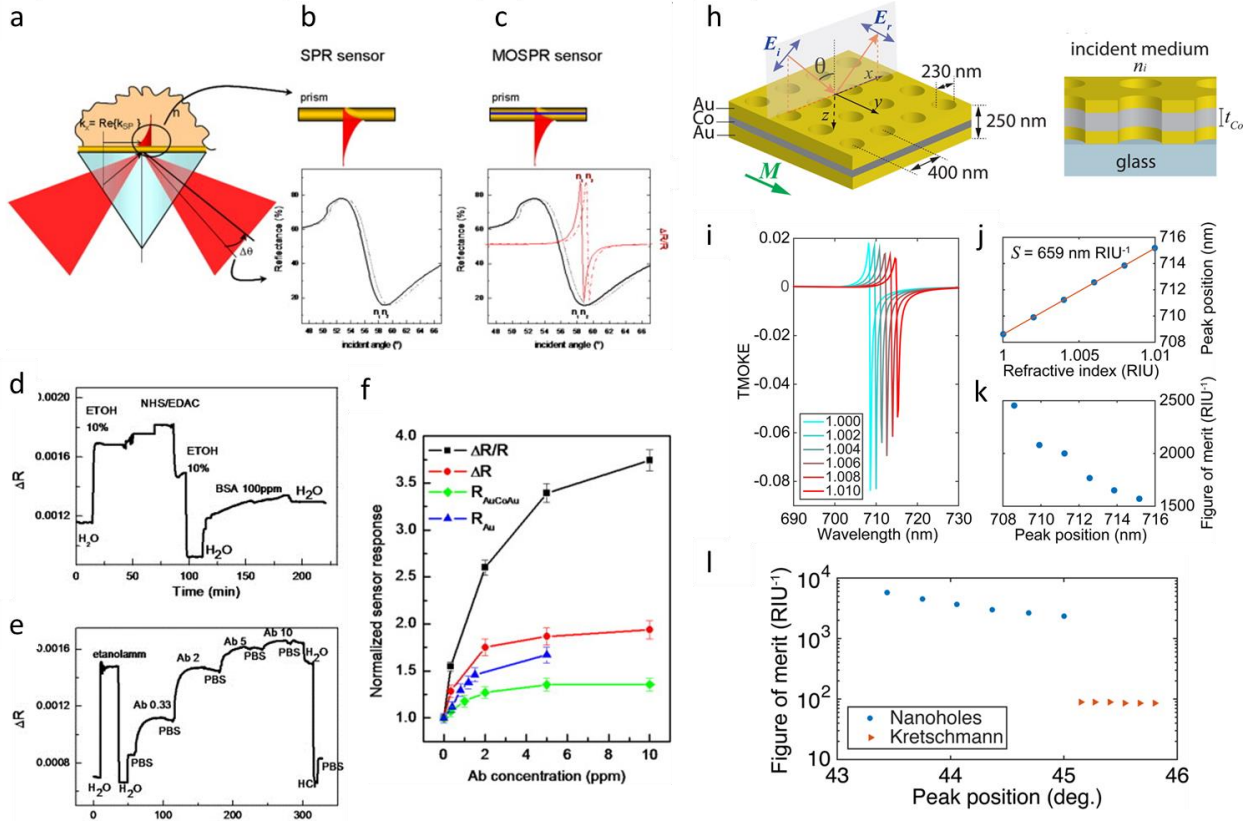

**Figure 2.** (a) Schematic representation of the experimental setup for SPP excitation in Kretschmann configuration; (b) detail of the metal transducer/dielectric interface of a standard SPP sensor and excitation of surface plasma wave exponentially decaying into the dielectric. The changes in SPP curves upon a change in refractive indexes $n_1<n_2$ is also shown; and (c) detail of the hybrid metal/ferromagnetic/metal trilayer transducer of a MO-SPP sensor and excitation of surface plasmon wave exponentially decaying into the dielectric. The EM field is enhanced at the MO active layer, thus enhancing the MO signal. A schematic response of SPP and MO-SPP signal to the same change in refractive index is reported. Dynamic ΔR curves in the MO-SPP sensor recording (d) the activation of the surface by NHS/EDAC solution and BSA immobilization after a proper washing in air, and (e) the binding with increasing concentration of anti-BSA antibodies (Ab) and regeneration step. (f) Calibration curves relative to antibodies (Ab) detection recorded by ΔR/R (squares), ΔR (circles), and reflectivity (rhombs) signals by using as optical and MO transducer the Au/Co/Au/Ti trilayer. A comparison with reflectivity sensor signal also obtained with a "classical" Au/Ti optical transducer (triangles) is reported. Reprinted with permission from Manera et al., Biosens. Bioelectron. **58**, 114 (2014). Copyright 2014 Elsevier. (h) Schematic representation of an Au−Co−Au perforated membrane with a periodic array of circular holes forming a square lattice, with the values of different geometrical parameters such as the lattice constant, the hole diameter, the membrane's total thickness, the Co thickness, the Co magnetization, M, which is parallel to the plane of the membrane and perpendicular to the incidence plane (left panel). Lateral view of the hybrid membrane that is placed on a glass substrate (right panel). (i) TMOKE of a Au−Co−Au nanohole array with a Co thickness of 110 nm as a function of the wavelength of the incident light for different values of the refractive index of the incident medium, $n_i$, which is assumed to be the same as that of the holes. (j) Position of the Fano-like feature as a function of $n_i$. The blue symbols correspond to the results of panel (i), and the red solid line corresponds to a fit to a straight line, whose slope defines the bulk sensitivity of our device, S = Δλ/Δni = 659 nm/RIU. (k) Corresponding figure of merit as a function of wavelength; see text for definition. (l) Comparison of the respective FoMs of sensors made of Au−Co−Au nanohole array with a Co thickness of 110 nm as a function of the angle of incidence and for a wavelength of 709 nm and of an Au−Co−Au planar trilayer (15 nm Au/6 nm Co/25 nm Au) in the Kretschmann configuration and for a wavelength of 632 nm. Reprinted with permission from Caballero et al., ACS Photonics 3, 203 (2016). Copyright 2016 American Chemical Society.

**IIb. Magnetoplasmonic sensing with LSPs**

An important part of the research in magneto-plasmonics is devoted to structures sustaining LSPs, such as nanoparticles. The rapid development of nanofabrication techniques allowed researchers to engineer complex plasmonic nanostructures, both substrate-supported and randomly dispersed in the investigated media. Molecular-level detection was reached with the simplest architecture, that is, nanodisk-shaped structures. The first proof of concept experiments of LSP magnetoplasmonic refractometric sensing was given by Bonanni et al. on a glass substrate [59] and by Pineider et at. with gold nanoparticles in solution [60]. Both groups showed that the MO response features a steep slope at the optical resonance, able to track RI changes of the surrounding medium. A significant improvement using pure FM nanodisks supporting LSP was reported in 2015 by Vavassori et al., who showed that Ni nanoantennas enable a raw surface sensitivity, without applying fitting procedures, of ~0.8 ag per nanoantenna of polyamide-6.6. Therefore, in principle it should be possible to detect sub-monolayers with polymers, peptides and proteins [61]. Their technique was based on the tracking of the phase of the LSP by measuring the polar MOKE signal and focusing on the ellipticity of the reflected/transmitted polarized light [62]. By varying the substrate, an additional phase can be added to the plasmonic one, and so the sensor can also work near topological darkness (that is pseudo- Brewster angle) condition [63,64]. Unlike amplitude-sensitive detection methods used in plasmonic sensing, phase-sensitive detection of MO signals provides stronger sensitivity for bulk and surface RI change monitoring [65]. In nanoplasmonic systems, the phase of the LSP mode defines the position of the optical resonance. Tuning the phase difference between optical and MO polarizabilities, achieved by exploiting the polarizability anisotropy in MP nanodisks geometry, can be used to exploit their MO activity for sensing [62]. A year later, Herreño-Fierro and co-workers studied Au/Co/Au MP nanodisks randomly arranged on a glass substrate and worked in TMOKE configuration upon Total Internal Reflection (TIR) excitation of LSP modes [66] (Figure 3a). By exploiting a similar phase-sensitive technique and focusing on the ellipsometric angle Δ working at the pseudo-Brewster angle condition, they obtained a MO surface sensitivity four times the typical sensitivity in SPP-based sensors (Figure 3b). Manera et al. examined this analytical aspect with a similar experimental approach but using bare metal nanostructures randomly arranged on a glass substrate [67]. First, they demonstrated numerically and experimentally a sizable MO activity in bare noble metal nanostructures in a TMOKE configuration upon modulation with a low-intensity magnetic field. When interrogated as RI sensing transducers, they resulted in a MO signal achieving a RI resolution of $5.6 \times 10^{-4}$ RIU, one order of magnitude higher than the respective LSP sensing signals. Overall, all these pioneering experiments with MP sensors supporting LSPs showed that these systems are ideal for achieving great FoM values compared to traditional plasmonic sensing platforms.

It is worth mentioning that different geometries showing a stronger plasmonic response, for instance, due to the hybridization of different plasmonic modes, might enhance the sensitivity performances of these systems. One type of architecture that might provide smaller effective volumes and a more robust MO response enabling even higher sensitivities is represented by the combination of a plasmonic nanoring and a magnetic disk [68,69]. This type of architecture, including split-ring resonators [70], may enhance the MO properties and display plasmonic modes beyond the usual electric dipolar LSP [71]. It would be interesting to understand their sensing capabilities and their detection performance limits. In this context, an interesting perspective comes from the exploitation of collective and/or coupling effects in magnetoplasmonic nanomaterials. Electromagnetically coupled structures have been reported by Dmitriev and co-workers to be very sensitive to the distance between the architecture building blocks. They have conceived a magnetoplasmonic dimer nanoantenna made of two Ni disks laying on the substrate and able to report nanoscale distances while optimizing its spatial orientation. The latter constitutes an active operation in which a dynamically optimized optical response per measured unit length allows for the measurement of small and large nanoscale distances with about two orders of magnitude higher precision compared to plasmon rulers made of noble metals [72]. Similarly, vertically coupled dimers have been proposed as an alternative [73,74]. Their MO response presents a Fano-like spectral lineshape, which might be exploited in sensing experiments, although no studies in this direction have been reported.

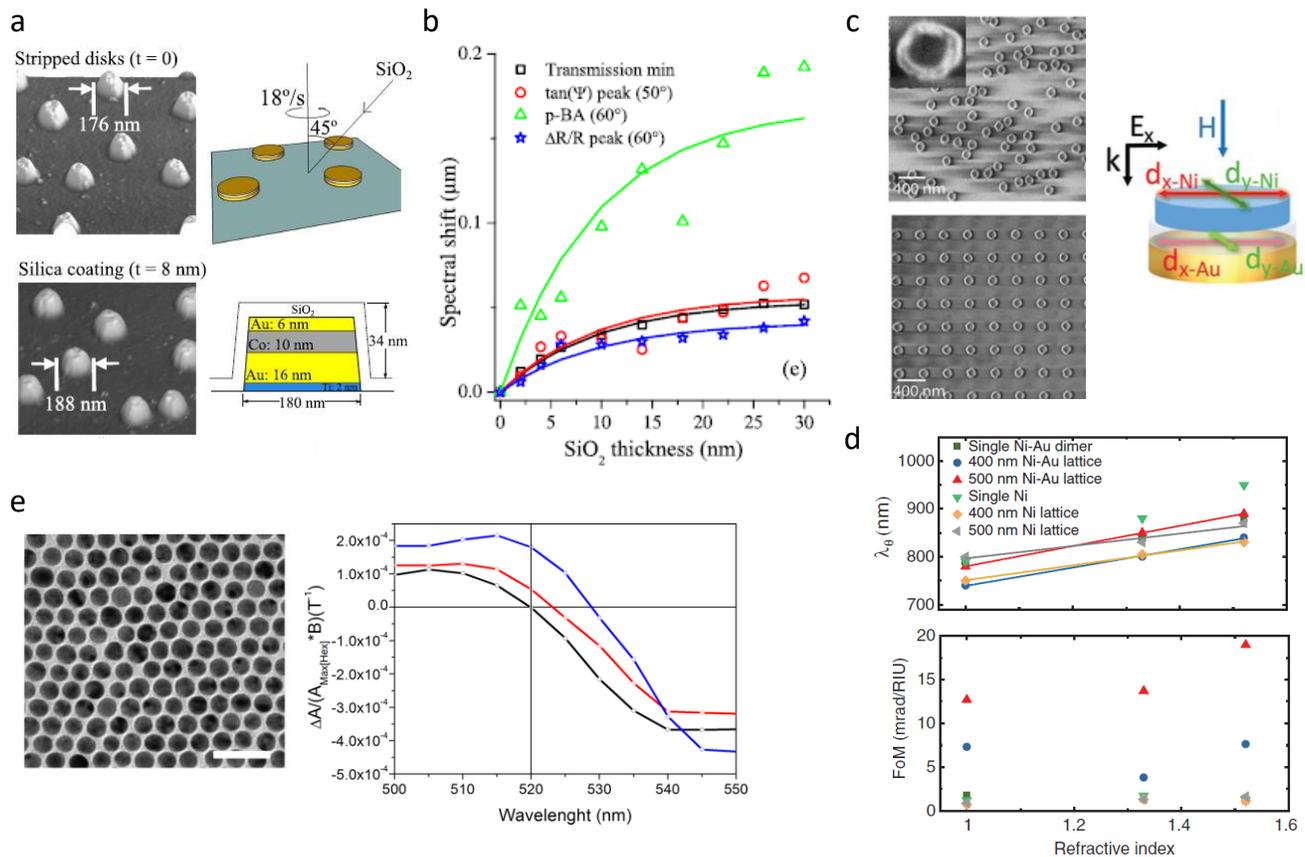

**Figure 3.** (a) 3D AFM images of nanodisks from stripped sample (t = 0 nm, top-left panel) and SiO$_2$ coated sample (t = 8 nm, bottom-left panel). Orientation of SiO$_2$ vapor flow 45° respect to z-direction grown at angular speed of 18°/s (top-right panel) to obtain coating layers with conformal shape as depicted in the bottom-right panel. (b) Spectral shift evolution of the studied structures. Solid lines correspond to the fitting as guide to the eye of the data. Reprinted with permission from Herreño-Fierro et al., Appl. Phys. Lett. 108, 021109 (2016). Copyright 2016 American Institute of Physics. (c) SEM images of samples with a random distribution (top-left panel) and a periodic (bottom-

left panel) array of Ni/SiO$_2$/Au dimers ($a$ = 400 nm). Right panel: schematic illustration of the dimer structure. The Au and Ni nanodisk are 30 nm thick, and their diameter is 150 nm and about 130 nm, respectively. The two metals are separated by 40 nm SiO2. Incident light with polarization along the $x$ axis excites electric dipoles in Au and Ni (*dx-Ni* and *dx-Au*) near the LSP condition. Spin-orbit coupling induces an orthogonal dipole in Ni (*dy-Ni*). Near-field coupling to this LSP in Ni also excites an orthogonal dipole in Au (*dy-Au*). (d) Summary of sensing performance of all magnetoplasmonic samples studied. Top panel: wavelength of zero Faraday rotation ($\lambda_\theta$). Lower panel: FoM as a function of the refractive index of the embedding medium. (e) TEM micrograph of colloidal gold nanoparticles; the scale bar is 40 nm. The average particle size is 12.8 ± 0.7 nm (left panel). Right panel: MCD spectra of 12.8 nm gold nanoparticles dissolved in hexane (black dots, RI = 1.375), in chloroform (red dots, RI = 1.446), and in toluene (blue dots, RI = 1.497). The vertical line at 520 nm highlights the significant changes in MCD signal magnitude as the RI of the medium is changed. Reprinted with permission from Pineider et al., Nano Letters 13, 4785 (2013). Copyright 2013 American Chemical Society.

To sharpen even more the resonant response of these systems, far-field coupled structures supporting surface lattice resonances (SLRs) have been also proposed [75,76]. In this case, the benefit is two-fold: thanks to the coherent interaction between the nanoantennas, the LSP and the MO signal are greatly enhanced, and the spectral features are sharper than those mentioned above, vertically coupled dimers. Thus, these nano-systems can be used for bio-sensing applications. On this topic, there is only a work focusing on RI sensing by van Dijken and colleagues, who introduced a system combining the benefits of both the near-field interaction in vertically coupled dimers and the far-field coupling giving rise to SLRs [77]. The sensor consists of a periodic array of Ni/SiO$_2$/Au dimer nanodisks (Figure 3c). Combined effects of near-field interactions between Ni and Au structures within the individual dimers and far-field diffractive coupling between the dimers of the array produce narrow linewidth features in the MO signal, which exhibit a spectral shift when the refractive index of the surrounding environment is varied. Because the resonances are sharp, refractive index changes are accurately detected by tracking the wavelength when the MO signal is close to zero. Compared to random distributions of pure Ni nanodisks or Ni/SiO$_2$/Au dimers, or periodic arrays of Ni nanodisks, the FoM of these hybrid magnetoplasmonic arrays is more than one order of magnitude larger [78] (Figure 3d).

A possible alternative to substrate-supported plasmonic nanostructures is the use of colloidal suspensions of plasmonic nanoparticles. These systems consist of metallic cores, surrounded by a layer of organic molecules, often referred to as surfactants, that stabilize the inorganic core against aggregation, ensuring at the same time favorable interactions with the solvent that allow the nanoparticles to form stable colloidal suspensions. Although not strictly following the definition of analytical transducers, since they are part of the solution volume, colloidal nanoparticles offer several advantages in refractometric sensing approaches. They can be produced in relatively large volumes with straightforward synthetic approaches, and can thus be regarded as disposable, floating transducers. Moreover, a great flexibility in materials design is granted by the chemical approach, as well as the possibility of preparing in parallel complex multi-component architectures. In a typical refractometric sensing experiment based on colloidal plasmonic nanoparticles, the shift of the LSP peak is monitored in transmission geometry as a function of the addition of analyte volumes. The extreme confinement of the electric field around individual nanostructures, granted by the small volume of the nanoparticles, further increases the selectivity toward surface adsorbed chemical species in comparison to larger, substrate-supported nanoantennas. Magnetic modulation of LSP has been demonstrated for simple colloidal gold nanoparticles: in a Magnetic Circular Dichroism (MCD) experiment, carried out in the Faraday geometry (magnetic field parallel to light propagation direction), the field-modulated signal of LSP shows a steep slope crossing zero in correspondence to the extinction maximum [60,79]. This is in principle a convenient detection scheme for high-performance refractometric sensing: in fact, the MCD signal exhibits substantial variations in intensity as a function of LSP peak position (Figure 3e). We can therefore define a performance index $Q$ for magnetoplasmonic refractometric sensing based on the MCD detection scheme as the variation of the MCD signal per change in refractive index $n$ around the nanostructure: $Q = \Delta MCD/\Delta n$ (given in units of ellipticity per refractive index unit, $\varepsilon/RIU$). The latter, in turn, depends on the ratio $Q_{MP} = \Delta MCD/\Delta \lambda_{LSPR}$ ($\varepsilon/RIU$), which is specific to the magnetoplasmonic response of the plasmonic transducer, while the dependency on the refractive index is described by the ratio $Q_{RI} = \Delta \lambda_{LSPR}/\Delta n$ (*nm/RIU*), which defines the refractive index sensitivity of a classical plasmonic sensing platform. It is then clear that the performance factor $Q$ is defined as $Q = Q_{MP} * Q_{RI}$. The optimization of $Q_{RI}$ has been the object of abundant literature on standard LSP-based sensing [80]. The term $Q_{MP}$, on the other hand, is the critical parameter that needs to be optimized for magnetoplasmonic refractometric sensing; in simple non-magnetic plasmonic systems, such as gold or silver nanoparticles, the Q factor can be as low as 0.03° $RIU^{-1}$, making the advantage of magnetic field modulation unsubstantial. To increase the magnetoplasmonic performance of colloidal nanosystems, attempts have been made to couple a magnetic moiety to the plasmonic one. Still, no interaction between the magnetic and plasmonic subunits could be observed [81,82]. On the other hand, small nanoparticles made up of an FM metal such as nickel do not show a sizeable LSP response due to the substantial overlap with interband transitions, which induce losses and damping of the resonance [83]. A promising route to increase the magnetoplasmonic response, and consequently the factor $Q_{MP}$ in colloidal systems, is the use of plasmonic heavily doped semiconductors [84]. In this class of semiconductors, degenerate doping causes a significant population of the conduction band, resulting in metallic systems able to support LSP in the infrared spectral region. The spectral position of the resonance can be tuned throughout the infrared through the choice of the host semiconductor and the amount of donor doping. An interesting feature for magnetoplasmonic refractometric sensing is that $Q_{MP}$ in non-magnetic systems scales with the cyclotron frequency, $\omega_c = eB/m^*$, where B is the applied magnetic field, and e and m* are the charges and effective mass of the charge carriers. While in metals, the value of m* is always close to that of the free electron $m_e$, in semiconductors it can be remarkably smaller. For instance, in indium tin oxide, the most known transparent conductive oxide, charge carriers have an effective mass down to $0.3m_e$. This, coupled to relatively sharp LSP, yields Q values of the order of 0.3 deg.$RIU^{-1}$ in the near-infrared, on par with the highest performance of FM nanostructures prepared by lithographic methods [85]. Building on these considerations, doped semiconductors based on InAs, with m*≈$0.02m_e$ would in principle afford even higher magnetoplasmonic sensing performance. This semiconductor, however, can only be doped to support LSP in the mid-infrared, thus making it not directly useful for refractometric sensing.

In conclusion, magnetic modulation of the LSP can be used to increase the sensing performance of colloidal nanostructures. The possibility of replacing noble and FM metals with heavily doped semiconductors seems promising in view of materials design and development for next-generation, high-performance magnetoplasmonic sensing. It is also worth mentioning an important work by Melnikau and co-workers, who further explored this topic by demonstrating the detection of strong MCD signal from supramolecular J-aggregates, a representative organic dye, upon binding to core−shell Au@Ag nanorods [86]. Here, the strong coupling occurring between exciton of the organic layer and NPs plasmon leads to the formation of a hybrid state in which the exciton effectively acquires magnetic properties from the plasmon. The above results are particularly exciting for sensing purposes but also pave the way toward selective enhancement of processes in close proximity to metal surfaces.

**III. All-dielectric nanostructures**

SPP sensors are a convenient and efficient tool for various biomedical measurements where the investigated biological objects, such as bacteria [87-90], are attached on the sensor's surface using one of the specific or non-specific adsorption methods, such as antibody immobilization and avidin–biotin interaction [87]. As already mentioned in the previous section, SPP sensors allow for the real-time study of various biological processes in the biological objects, including the response to the medical treatment and action of antimicrobial drugs in vitro, in particular [91]. However, the main disadvantage of SPP based sensors is the low Q-factor which is the relation of the SPP wavelength to the SPP spectral width. SPP Q-factor usually is Q~10 due to the high losses in metals, and such low values significantly limit SPP sensor sensitivity. Another issue for the biosensing is the relatively small (about 100 nm) penetration depth of the EM field into the analyzed substance that is significantly smaller than typical bacteria sizes of about 1 µm. Also, SPP sensor signal includes any possible fluctuations of the refractive index of the analyte such as variations of the mixture composition, temperature, etc. To extract the signal that is associated with the analyzed biological objects, a reference cell that measures the bulk refractive index of the solution is often used. In this respect, all-dielectric structures for sensing are more advantageous since they possess sharp resonances and allow to detect simultaneously in several channels responsible for the studied object and surrounding medium, correspondingly, e.g., using p- and s-polarized illumination. Here we will focus our attention on the planar multilayer structures bearing in mind the simplicity of their fabrication and utilization in the sensing process. This section describes how the all-dielectric magnetophotonic crystals could help to overcome the problems mentioned above.

**IIIa. Tamm surface modes in photonic crystals**

The most straightforward way to increase the sensor detection level is to substitute the metallic layers with the dielectric ones in order to sharpen the optical resonance, which broadens due to the high losses in metals. However, while the plasmonic metal/dielectric structures provide the possibility to excite surface waves (SPPs, namely), at the interface between two dielectrics, the existence of such waves is forbidden in the case of isotropic materials [92].

Although surface waves, named Dyakonov waves, could be excited in the presence of the optically anisotropic crystals, it is hard to benefit from this configuration for sensing. The reason is that the Dyakonov waves are excited under the awkward conditions $\varepsilon_o < \varepsilon_{an} < \varepsilon_e$ where $\varepsilon_{o,e}$ are the ordinary and extraordinary permittivities of the crystal and $\varepsilon_{an}$ is the permittivity of the analyte [92], which is $n \sim 1 \ldots 1.4$ for the typically analyzed substances. The other option would be to use guided modes in a high refractive index thin film since they also produce an evanescent field in the analyte. Let us consider a simple planar waveguide structure: prism/cladding/waveguide/analyte, where the role of the second cladding is taken by the analyte. The dispersion of the guided waves is sensitive to the refractive index of the analyte. However, this contribution is accompanied with the influence of the core layer and the other cladding. For the efficient excitation of the guided waves, one needs to have a high contrast between the core and the cladding refractive indices, but it significantly diminishes the effect of the change of the analyte refractive index on the guided mode dispersion. The sensitivity of the guided mode grows near the guided mode cut-off conditions where the mode field concentrates near the waveguide surface whose refractive index is close to mode one. Nevertheless, typically it is not possible to find a dielectric material for the waveguide cladding with the refractive index lower than that of the analyte. As a result, the cut-off conditions with the mode shifted to the analyte surface remain impracticable.

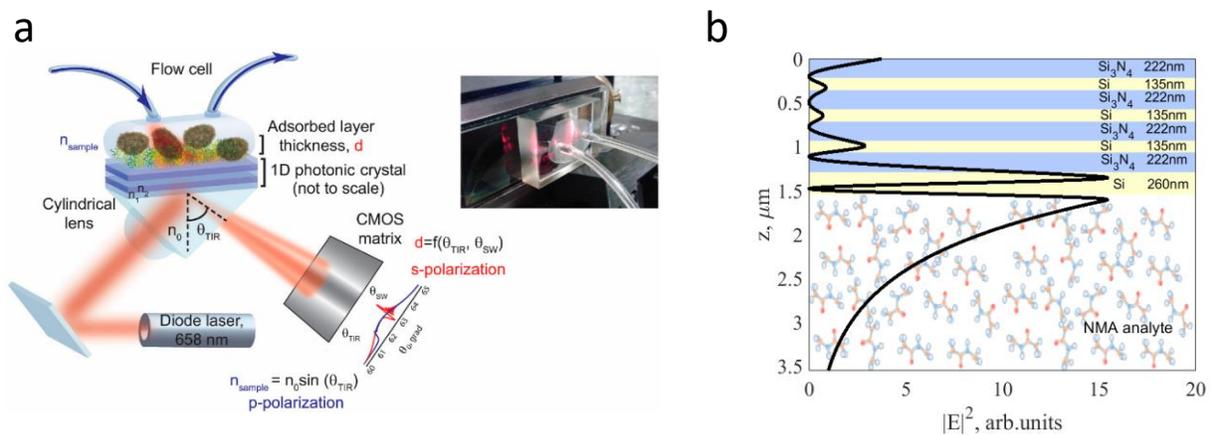

**Figure 4**. Biosensing with the Tamm surface modes. (a) Principal sensor scheme. Reprinted with permission from Rostova et al., Biosensors 6, 52 (2016) [93]. Copyright 2016 MDPI. (b) The Tamm surface mode profile in the PC-based structure. Reprinted with permission from Borovkova et al., Photonics Res. 8, 57 (2020) [94]. Copyright 2020 The Optical Society.

In the best realistic scheme utilizing a prism-SiO$_2$($n_{SiO2}$=1.45)-Ta$_2$O$_5$ ($n_{Ta2O5}$=2.0, $h_{Ta2O5}$=50 nm) as the sensing structure and water as an analyte, one gets the sensitivity of $\partial n_{GM}/\partial n_{an} \sim 0.03$. A simple SPP-based structure with an Au layer gives a sensitivity of $\partial n_{SPP}/\partial n_{an} \sim 1$. Therefore, although one may benefit from the high Q-factor and long-range-propagating guided modes emerging in all-dielectric structures, the sensitivity of the prism-waveguide schemes is still lower or comparable to the plasmonic ones (see, for example, Ref. [95]). The properties of a multilayer medium consisting of the specially-arranged hundred-nanometer thick dielectric components significantly differ from that of bulk materials. It was shown that the interface between the two all-dielectric PCs could support surface EM mode [96,97], so-called Tamm surface mode (TSM). Similar modes were observed at the surface of a single PC under the total internal reflection conditions from the bounding medium (Figure 4a,b). Such modes are bounded due to the bandgap of PC and total internal reflection. First, such surface states were predicted for electrons at the surface of the crystals by I.E. Tamm [98].

There are several advantages of the Tamm modes comparing to the SPP modes. First, both TM and TE modes can be excited since the bandgap of the PC can be realized for both light polarizations. Second, as the structures can be made from the dielectric material, the absorption losses are nearly zero, and therefore Tamm modes possess ultrahigh Q-factors that are 2-3 orders higher than the Q-factors of SPP. Finally, the dispersion of the Tamm modes can be tuned 'by design', i.e. by selecting the PC parameters, such as layer thicknesses and materials, and a number of layers. This offers the possibility of tuning the detection area of the sensors, as will be discussed later. Moreover, one may control the frequency of the Tamm mode and excite it even in the ultraviolet spectral range [99] where the conventional SPPs are not allowed due to the dispersion of metal permittivity. Tamm modes in the UV range are significant for fluorescence biosensing [100].

Actually, one does not need to introduce the metal layers in PC to excite the Tamm mode. However, if the surface of PC is covered with a metal layer, so-called Tamm plasmons can be excited, which are the hybrids of the Tamm modes and surface plasmon-polaritons [101]. They also possess high Q-factors and ultralong propagation lengths of up to millimeter distances [102]. One may benefit from the excitation of Tamm plasmons instead of Tamm surface modes in the cases where the presence of metal is unavoidable for sensing, for example, hydrogen sensing using Pd layers [103,104] or nitrogen dioxide sensing using Au layers and UV range [105].

**IIIb. Magneto-optics of Tamm surface modes**

To understand how the combination of the Tamm surface modes with the MO materials can improve the sensing process, let us briefly discuss how the optical response of TSM is modified in the presence of the external magnetic field applied to the PC structure. To exclude the additional absorption losses, the introduction of the transparent dielectric materials in TSM-supporting PC, such as iron-garnets, is preferable [106, 107]. If a PC possesses MO properties, surface modes at its interface modify. Actually, under the application of the magnetic field in the transverse configuration, the propagation constant of the Tamm mode $\beta^{\text{TM}}$ is changed [108,109]:

$$\beta^{\text{TM}} = \beta_0 \pm \Delta\beta_H,$$

where the sign before $\Delta\beta_H$ depends on the direction of the magnetic field, and the modulation is observed only for the TM-type mode. This results in the shift of the TSM resonance position, which leads to the modulation of the reflectance coefficient (TMOKE). For the two other configurations of the external magnetic field, the longitudinal and polar, respectively, the propagation constant depends on the medium magnetization quadratically, and therefore, it is possible to neglect it. On the other hand, the polarizations of the two eigenmodes modify linearly with magnetization [110]:

$$\Psi^{q-TM} = \Psi^{TM} \pm \alpha_1 \Psi^{TE},$$

$$\Psi^{q-TE} = \Psi^{TE} \pm \alpha_2 \Psi^{TM},$$

where $\Psi^j$ are the polarizations of the TM, TE modes in the non-magnetic case and the mixed quasi-TM (q-TM) and quasi-TE (q-TE) modes in the magnetized structure, and $\alpha_{1,2}$ are the coefficients proportional to the material gyration. Such polarization modification of the Tamm modes reveals itself in the two types of MO effects. First, the Faraday rotation is enhanced due to the eigenmode excitation comparing with a magnetic film of the same thickness without a PC [111-113] and this effect is odd in the magnetization of the structure. The second effect arises from the change of the coupling efficiency of the incident light to the TSM, which is proportional to $\cos^2(\Psi^{inc} - \Psi^{TSM})$, where $\Psi^{inc}$ is the polarization of the incident light and $\Psi^{TSM}$ is the polarization of the TSM. This effect is called a magnetophotonic intensity effect and was demonstrated for the longitudinal [114] and polar [115] configurations of the magnetic field and can also be observed in specially designed structures in a transverse magnetization configuration [116]. Due to the mixed eigenmode polarization, the resonance of TSM is seen as the intensity effect in both polarizations of the incident light [117].

**IIIc Sensing with Tamm surface modes in magnetophotonic crystals**

Magneto-optics could improve the performance of TSM-based sensors. The most obvious advantage is to increase the sensitivity of the sensor if the MO modulation $\delta$ of the reflectance $R$ under the application of the external magnetic field

$$\delta = \frac{R(H_1) - R(H_2)}{R(H_1) + R(H_2)}$$

is measured instead of the reflectance measurements. In the simplest case, it is a TMOKE effect with $H_1 = +H_y$ and $H_2 = -H_y$. Both TMOKE and reflectance resonances are associated with the TSM so that they experience the same shift in the angular or wavelength spectra under the variation of the refractive index of the analyte (Figure 5a,b). However, the signal-to-noise ratio of the TMOKE signal is higher than that of the reflectance. Although in the transparent magnetic materials without any modes, TMOKE has a relatively low value, excitation of the TSM allows for its increase up to $\delta = 10 - 20\ \%$ [46,118]. The angular spectra of $\delta$ have a very high derivative $\partial\delta/\partial\theta$, and $\partial\delta/\partial\theta$ is largest exactly at the TSM resonance. The latter favors the sensitivity since at the TSM resonance maximal part of the incident energy is coupled to the optical mode. This is in contrast to the $R$ spectrum that has a zero derivative at the TSM resonance, and one has to work at the resonance slope where the coupling to the optical mode is less efficient. This makes the sensitivity of the MO sensing $\partial\delta/\partial n$ to be much larger than that of the purely optical one $\partial R/\partial n$ (Figure 5c). On the other hand, the noise in the TMOKE measurements is lower since $\delta$ is the relative variation of the reflectance, the presence of some spurious fluctuations of the laser intensity, interference, etc. makes very small impact on it. Both the increase of the signal and reduction of the noise result in a significant, almost 2 orders of magnitude larger, MO sensor sensitivity and detection limit compared to purely optical measurements [46,118].

TMOKE is extremely sensitive to light absorption [19]. The optical losses in the iron-garnets [107], as well as in the dielectric layers of the PC, could be very low if near infra-red spectral range is used. This makes the situation different from that with the SPP structures, where the losses due to the high metal absorption are unavoidable. The absence of the absorption in MPC leads to the sensitivity of the TSM resonance width and depth to the imaginary part of the refractive index of the analyte. Reflectance resonance could be described by a Lorentzian shape:

$$R = 1 - \frac{\Gamma_i \Gamma_{rad}}{(k_0 \sin\theta - \beta)^2 + (\Gamma_i + \Gamma_{rad})^2},$$

where $k_0$ is the wavevector in a vacuum, $\theta$ is the angle of incidence and $\Gamma_i$ and $\Gamma_{rad}$ are the TSM losses due to absorption (i) and radiation leakage (rad).

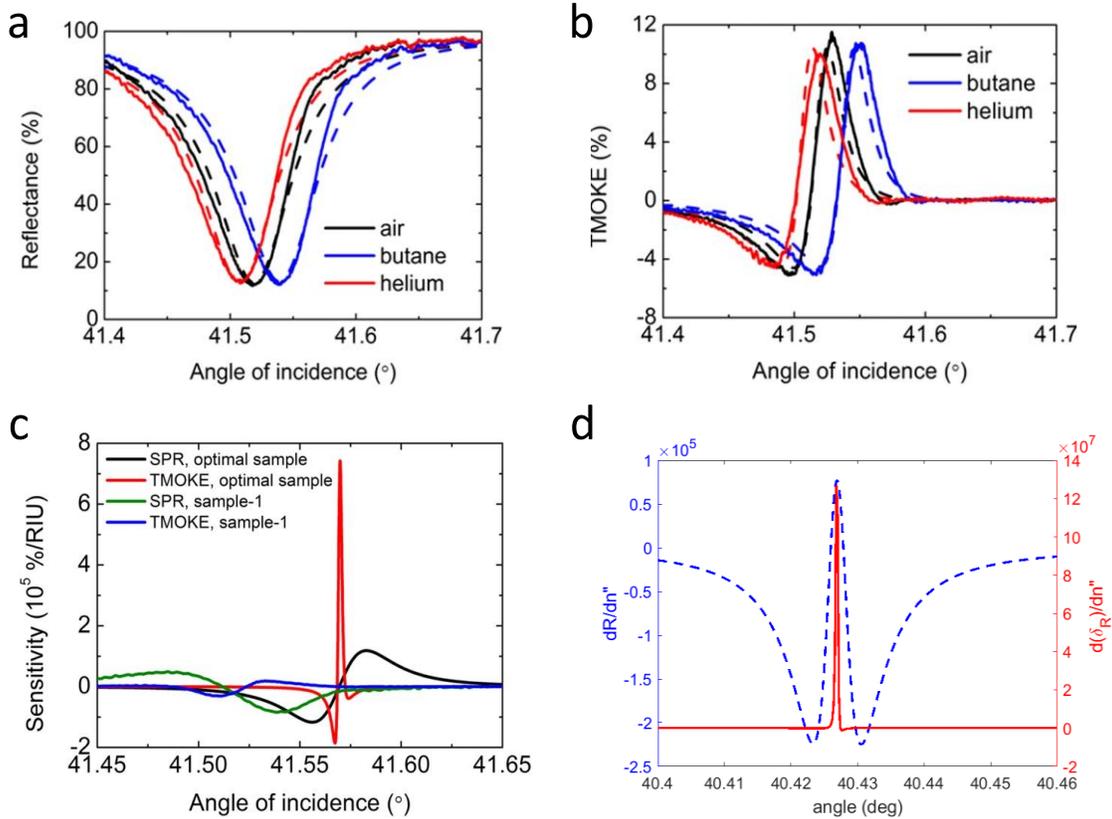

**Figure 5.** Optical and MO sensing with the Tamm surface modes. (a) Reflectance spectra and (b) TMOKE spectra of the same MPC bounding different gases (see the legend). (c),(d) Comparison of the optical and MO sensitivity to the variation of (c) Re(n) and (d) Im(n) in MPC with Tamm surface modes. Images (a)-(c) are reprinted with permission from Ignatyeva et al., Sci. Rep. 6, 28077 (2016). Copyright 2016 Springer-Nature. Image (d) is reprinted with permission from Borovkova et al., Photonics Res. 8, 57 (2020). Copyright 2020 The Optical Society.

The presence of small absorption peaks that can arise, for example, in the near-infrared region due to the molecular vibrational overtones [119], will modify the $\Gamma_i$ significantly if it is the only source of the absorption in the system, Thus one may obtain the sensitivity of the TSM resonance to the imaginary part of the refractive index several orders exceeding the sensitivity of the similar structure without TSM [94,120]. Here again, even higher sensitivity could be obtained in measuring the TMOKE spectra (Figure 2d) [94]. The advantages of the MO approach are the same as for sensing the real part of the refractive index discussed above with a sensitivity increase up to 2 orders of magnitude [94]. However, it is essential to underline that such sensing itself is possible only due to the nearly zero absorption in the structure supporting the surface mode.

Finally, there is an important point on the detection area $L$ which is defined as the penetration depth of optical radiation into the analyzed substance. $L$ can be determined using the following relation:

$$L = \frac{\lambda}{4\pi\sqrt{(n_{prism}\sin\theta)^2 - n_{an}^2}},$$

where $\lambda$ is the light wavelength, $n_{prism}$ and $n_{an}$ are the refractive indices of the coupling prism and the analyte, respectively. $L$ tends to infinity at the total internal reflection (TIR) angle and reaches ~1 µm in its vicinity. However, for the angles that are several and more degrees higher than the TIR angle, $L$ is almost constant and is around 0.1-0.2 µm. It constrains measuring large biological objects, such as bacteria, etc., with the conventional SPP sensors that, due to the wide SPP widths of 5-15 deg., cannot operate close to the TIR. The width of the TSM resonance could be several orders of magnitude smaller, up to 0.01 deg, and varying the parameters of the PC, one can tune the $\theta$ angle to get almost any desired penetration depth, from almost infinite near TIR angle to 0.1 µm (Figure 6c).

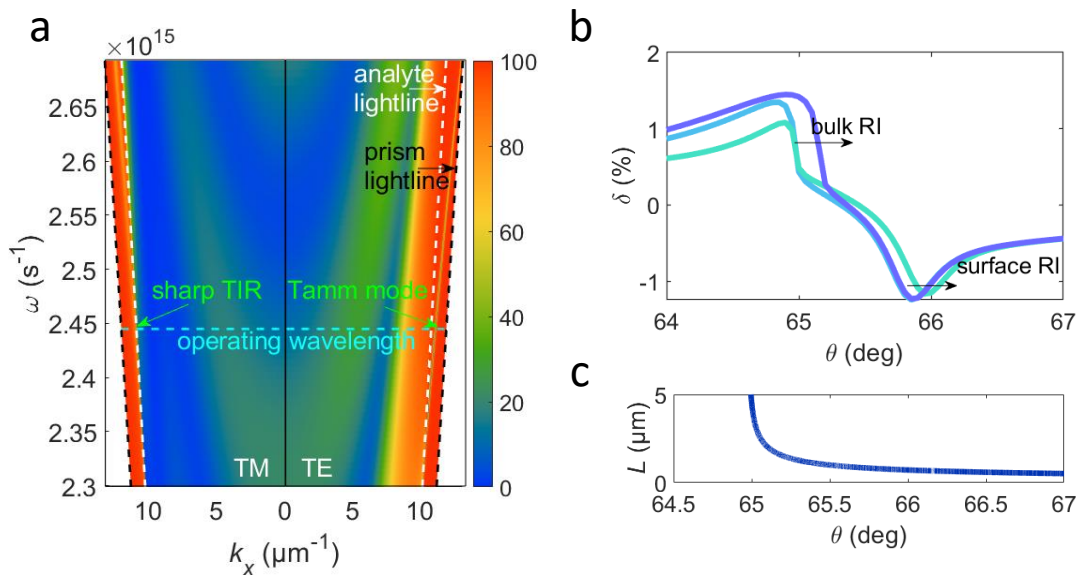

**Figure 6.** Sensing with Tamm surface modes. (a) Reflectance spectra of PC with Tamm surface mode in TE and sharp TIR in TM. (b) Spectra of MO response with simultaneous sensitivity to the bulk and surface refractive indices (c) The dependence of the radiation penetration depth on the position of the resonance. Re-adapted with permission from D. Ignatyeva et al., Sensors 21, 1984 (2021Copyright 2021 MDPI.

Actually, one can use both TM and TE polarizations in the frame of a single device [93] (Figure 6a). They have different dispersion and, therefore, penetration depth so that it is possible to obtain data on the changes in the surface and bulk characteristics of the analyte in a dual-channel mode in p- and s- polarized light, correspondingly [93,121]. The MO scheme of measurements could improve the performance of such sensor since both of the TM and TE modes are observed in the same $\delta$ spectra if both $H_1$ and $H_2$ have the transverse and polar components [115]. In this case, two resonances in the $\delta$ spectra of the same beam are independently sensitive to the bulk and surface refractive indices (Figure 6b).

Therefore, the Tamm surface mode-based magnetophotonic all-dielectric crystals advance the sensing process. They possess sharp MO resonances making the sensing more precious compared to both magnetoplasmonic and optical TSM-based sensing. Elimination of the absorbing metal layers and measuring the absorption-dependent MOKE allow achieving high sensitivity to the small absorption peaks in the studied materials. At the same time, the penetration depth of the Tamm surface mode can be controlled by tuning the parameters of the PCs and, in contrast to the highly-confined SPP, can have the same order as the measured large, micron-sized, biological objects. It is possible to design the PC supporting simultaneously Tamm modes in the two orthogonal polarizations that have significantly different penetration depths. Application of the external magnetic field provides both resonances in the MO signal measured for the fixed polarization of light. This allows one to distinguish a signal responsible for the studied object and a signal responsible for the variations of the surrounding medium.

**IV. Hybrid and novel magnetoplasmonic materials**

From the previous sections, we can infer that all-dielectric magnetophotonic structures can partly overcome the limitations associated with conductive losses from metallic materials [122-125], but magnetoplasmonic structures offer higher MO activity with nanostructures having dimensions smaller than the incident wavelength [126-130]. Hybrid nanostructures comprising noble metals and dielectric magnetic garnets allow for the trading-off of these respective advantages, enabling high MO activity with reduced levels of losses, without losing the ability to confine and enhance EM fields at subwavelength regions [131-135]. Since noble metals exhibit smaller losses than FM metals, hybrid noble-FM metallic nanostructures have also been exploited for magnetoplasmonic approaches [46,53,77,136,137]. These works have motivated various applications, including for sensing [51,61,78,115,138-139], telecommunications [140-144] and magnetometry [145,146]. It is significant that the large losses from hybrid magnetoplasmonic nanostructures, considered a major drawback for biosensing, are suitable for nanoparticle-based hyperthermia and drug delivery [147-150]. The EM energy absorbed by the nanoparticles can be used for thermal ablation of dangerous cells in the human body [12,147,149] and for drug delivery on demand by thermal activation [148] (or by breaking thermally sensitive bonds [150]), respectively.

Despite the achievements of the last two decades, there is still intense research activity on magnetophotonics, and magnetoplasmonics fueled not only by the need to diminish the losses but also enable the active control of nanophotonic devices. Indeed, externally applied magnetic fields can be used to control the optical chirality of magnetoplasmonic surfaces [151-153]. Significantly, the combination of magnetically controlled optical forces [154-156] and switchable optical chirality in magnetoplasmonic surfaces [151-153] is promising for enantioselective optical trapping and sensing chiral samples up to the single-molecule level. For example, the switching of extrinsic optical chirality by applied magnetic fields was recently demonstrated using a magnetoplasmonic device [153]. A schematic representation of the device, comprising a gold thin film perforated with a hexagonal periodic array of nanoholes, fabricated on a Ce:YIG/YIG/SiO$_2$/TiN multilayer and deposited on a SiO$_2$ substrate, is shown in Figure 7(a). A cross-sectional transmission electron microscopy (TEM) image of the nanodevice is shown in Figure 7(b). This structure enables the tuning of the extrinsic chirality exhibited by the nanohole array by varying the magnitude and sense (along the $z$-axis) of the applied magnetic field, as observed in the experimental results in Figure 7(c) [153]. Under oblique incidence without magnetic field (W/O H), the structure displays angle-dependent circular dichroism (CD) spectra due to the excitation of asymmetric cavity modes for right (RCP) and left (LCP) circularly polarized light. If an external magnetic field is applied perpendicular to the plane of the structure, the chiroptical activity can be continuously tuned and even switched, as observed from this last figure.

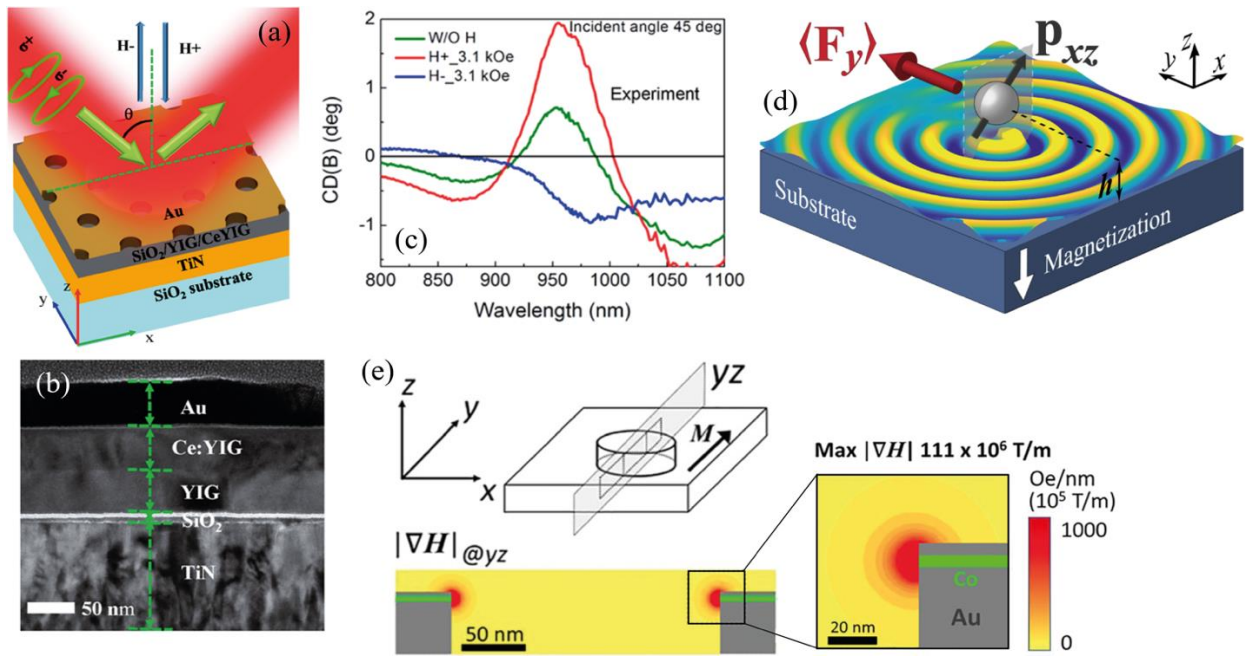

**Figure 7.** (a) Pictorial representation of the device and its operation mechanism. **H**+ and **H**− indicate the sense ± of the applied magnetic field along the $z$-axis. $\sigma$+ and $\sigma$− are used to indicate the RCP and LCP incident light, respectively. (b) Cross-sectional TEM image to show the individual layer thicknesses. (c) CD measured for an incident angle $\theta = 45°$, as a function of the incident wavelength, for magnetic field amplitudes varying from -3.1 kOe (oriented along the +$z$ direction) to +3.1 kOe (oriented along the −$z$-direction). Reprinted with permission from Qin et al., ACS Nano 14, 2808 (2020). Copyright 2020 American Chemical Society. (d) Schematic of a dipolar particle **p**$_{xz}$ linearly polarized along the $xz$-plane, placed at a distance $h$ above a MO substrate. The magnetization of the substrate is considered along the $z$-axis. Above the substrate is shown the radiated Re{$H_z$} field. Fields reflected from the surfaces undergo polarization conversion introducing

directionality of emission along the *xy*-plane and a corresponding lateral optical force. $<F_y>$ is illustrated for the corresponding configuration of magnetization and polarization. Adapted with permission from Girón-Sedas et al., Phys. Rev. B 100, 075419 (2019). Copyright 2019 American Physical Society. (e) The upper panel on the left side sketch the *yz*-plane along the center of a magnetoplasmonic nanopore, parallel to the magnetization (**M**) of the MO Co layer. The lower panel on the left side shows the magnetic field gradient, $\nabla \mathbf{H}$, calculated along the *yz*-plane. The right panel is a zoom to illustrate the amplitude of the corresponding magnetic field gradient. Reprinted with permission from Maccaferri et al., Appl. Phys. Lett. 118, 193102 (2021). Copyright 2021 American Institute of Physics.

This phenomenon can be explained by the active control of the electric near-field amplitude inside the hole due to the plasmon-enhanced (Au/Ce:YIG) MO activity of Ce:YIG. Moreover, this configuration for the magnetization, i.e., normal to the surface, was demonstrated to produce lateral optical forces due to reflected field polarization conversion effects, as illustrated in Figure 7(d) [155]. The direction and magnitude of this lateral force will depend on the magnetization sense and the polarization of the nanoparticle, as it was shown for linearly and circularly polarized particles. Effective approaches for magnetoplasmonic enantioselective chiral trapping (separation) and sensing, even up to the single-molecule level, can be envisaged by merging these last concepts, which is of prime importance for the pharmaceutical industry [157]. On the other hand, for the in-plane magnetization configuration, improved gradient electric [154] and magnetic [156] optical forces have been demonstrated. The top panel on the left side of Figure 7(e) illustrates an Au/Co/Au nanopore, with the in-plane magnetization pointing along the *y*-axis. The modulus of the magnetic field gradient along the *yz*-plane is shown in the lower panel, reaching values up to $10^8$ T/m at the edge of the nanopore, which can be exploited for trapping of magnetic and magnetoplasmonic nanoparticles [156].

Magnetoplasmonic metamaterials are also being extensively investigated. In contrast to conventional grating- and waveguide-based magnetoplasmonic structures, metamaterials work in the effective medium regime, thus being robust to small-scale fabrication errors. Moreover, the *ε*-near-zero (ENZ) mode from metamaterials or semiconducting-oxide materials can be used to overcome the need to use prisms or grating couplers for SPP-based magnetoplasmonic sensing, as shown with numerical simulations and experiments [158-161]. This latter approach exploits the coupling between highly absorbing ENZ and SPP modes at hybrid magnetoplasmonic interfaces. Other recent strategies consider hybrid uniaxial MO metamaterials fabricated through magnetoplasmonic core/shell nanopillars [162,163], oxide-nitride magnetoplasmonic nanorods [164] and multilayered metal-dielectric hyperbolic nanoparticles with strong optical anisotropy [165-167] for developing tunable hybrid materials integrating plasmonic, magnetic and dielectric properties. However, the MO response from these metamaterials is still considered low to reach applications, thus requiring further research in terms of materials and methods.

A promising class of materials for magnetoplasmonics are heavily doped semiconductors, which are able to support plasmonic excitations over the whole infrared range, which can be tuned with the amount of doping [84]. As discussed in section IIb, LSP in heavily doped semiconductors can be very sharp even compared to state-of-the-art plasmonic metals such as gold and silver. In addition, a low carrier effective mass affords an exceptionally strong response to magnetic field even in purely diamagnetic systems [85]. Finally, the possibility of co-doping with magnetic ions paves the way for potential game-changer materials in the field of magnetoplasmonics and magnetoplasmonic sensing.

## V. Magnetoplasmonic magnetometry

It is natural to use the MO effects in nanostructures for magnetometry which is sensing of the magnetic field itself. Magnetometers are important, for example, for various biomedical technologies, such as magnetoencephalography and magnetocardiography, which require the measurements of superweak magnetic fields (of the orders of nano- and pico-Tesla) [168]. Various magnetometers are commercially available [169], with SQUID (Superconducting Quantum Interference Device) being the leader with a sensitivity of up to 1 fT. The field of MO magnetometers is actively developing whose advantages are the simplicity of operation (cryogenic temperatures are not needed), the ability to obtain a spatial resolution of the order of 10-100 nm in a direction orthogonal to the magnetic film, and the ability to scan with a laser beam and obtain micron resolution in the lateral direction [170]. It is important that MO magnetometry allows one to perform measurements in vector format [171]. The main disadvantage of MO magnetometers is the significant loss in sensitivity compared to SQUID magnetometers, by 2-3 orders of magnitude. The sensitivity of MO magnetometers depends on the magnitude of MO effects in the structures, as well as on the magnitude of the base signal (that is, the transmittance or reflectance of the structure). The application of various nanostructured materials is essential to improve the signal-to-noise level. The photonic-crystal based microcavity with a magnetic defect is known to provide multifold enhancement of the Faraday effect, therefore enhancing the sensitivity to the normal magnetic field component [172]. The sensitivity to the in-plane constant (DC) transverse components of the external magnetic field down to 0.3 nT (for 1 mm$^2$ area) was obtained in the measurements of the hysteresis loops of the magnetoplasmonic crystals [173] which exhibit the uniquely high TMOKE [174].

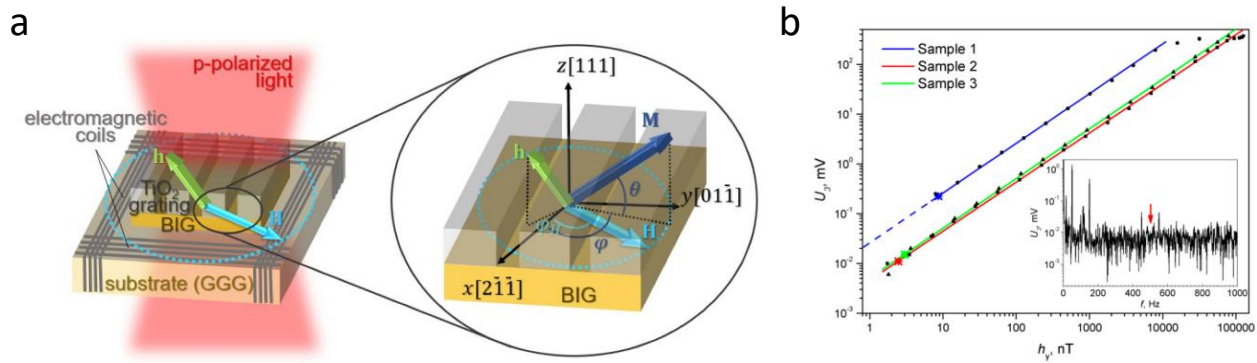

**Figure 8.** 1D grating-based MO vector flux-gate magnetometer. (a) Device principal scheme. The inset shows magnetization M motion in the iron garnet film under the rotating external magnetic field H and the measured weak magnetic field h. Reprinted with permission from Ignatyeva et al., J. Phys. D: Appl. Phys. 54, 295001 (2021). Copyright 2021 American Institute of Physics. (b) Dependence of the MO signal U3, on the monitored magnetic field hy oscillating at 515 Hz (indicated by a red arrow on the noise characteristic in the inset). Inset: Noise characteristic of the magnetometer setup). Reprinted with permission from Knyazev et al., ACS Photonics 5, 4951 (2018). Copyright 2018 American Chemical Society.

More advanced is the fluxgate technique, in which an external magnetic field exceeding the saturation field rotates in the plane of the magnetic film [175], modulating the intensity of light in a nonlinear asymmetric manner due to the presence of the measured weak in-plane components. This technique makes it possible to find simultaneously two spatial components of the external magnetic field in the plane of the sample down to ~ 1nT magnitude both with an ultrahigh sensitivity of $fT/Hz^{1/2}$ and micron spatial resolution [176]. Full vector magnetometry with all 3 spatial components of the external magnetic field can be measured using the MO Faraday effect in films with a special type of anisotropy. In such anisotropic magnetic films, the rotation of the external magnetic field in the plane leads to the appearance of both lateral and normal magnetization components, each of which depends in a complex way on the direction of the external magnetic field. A series of frequency harmonics appear in the temporal spectra of the observed Faraday effect, with the amplitude corresponding to the measured weak magnetic field components down to 0.1 nT magnitudes each [177]. The utilization of the nanostructured iron-garnets (Figure 8) with the guided modes, possessing both high transparency and large MO response, helps to increase the sensitivity and spatial resolution of such magnetometers [146].

**VI. Towards industrial applications**

Sensors are now ubiquitous, being used from electronic to mechanical and magneto-plasmonics. Magneto-plasmonic sensors have attracted interest in the industry due to the possibility of improving detection speed, sensitivity, and portability [178]. In addition to supporting surface plasmon resonance phenomena, there exists a possibility of combining and optimizing these nanostructures in various combinations (from multilayer to nanoparticles), compositions (0 to 100 at. %), thicknesses (0 to 100 nm), and magnetic activity (H = 0 to 1 kOe) to tune and significantly enhance the sensitivity and detection limit of the sensor. However, the industrial applications of these sensors are still far from reality as they require a minimum of nine stages from concept to commercializing, as identified by industry Canada. The preliminary stage involves a thought process and gathering information about the sensor. Then comes the question of translating ideas into applied research. In this stage, one can observe and report basic principles of the concept, going through literature reviews to conducting analytic studies for practical applications, followed by experimentation of individual components such as testing and validation in the laboratory. The next stage involves integrating and developing the product, testing and incorporating electronic, optical, microfluidics, mechanical hardware, and software in the laboratory setting, leading to a model or prototype in a simulated operational environment. The final stages involve a prototype ready for demonstration and developmental testing. Through many iterations, the actual application of the technology takes place in its final form and under real-life conditions.

As discussed in the previous sections, magnetophotonic sensors involve a combination of dielectric and magnetic materials in various compositions and layer thicknesses. Significant work has been carried out in research laboratories worldwide, demonstrating sensitivity tuning at both visible and near-IR ranges by manipulating plasmonic behavior with magnetics [179]. Compared to the plasmonic behavior shown by the conventional surface plasmon resonance-based devices [180], magneto-plasmonic configurations offer a sharper slope of the responses produced by an externally modulated magnetic field. They also allow amplified sensitivity and improved detection limit, as high as 2 x $10^{-8}$ RIU for bilayer or bilayer/multilayers when combined with a PC [46,181]. Thus, it offers many adjustments possible to optimize sensor performance.

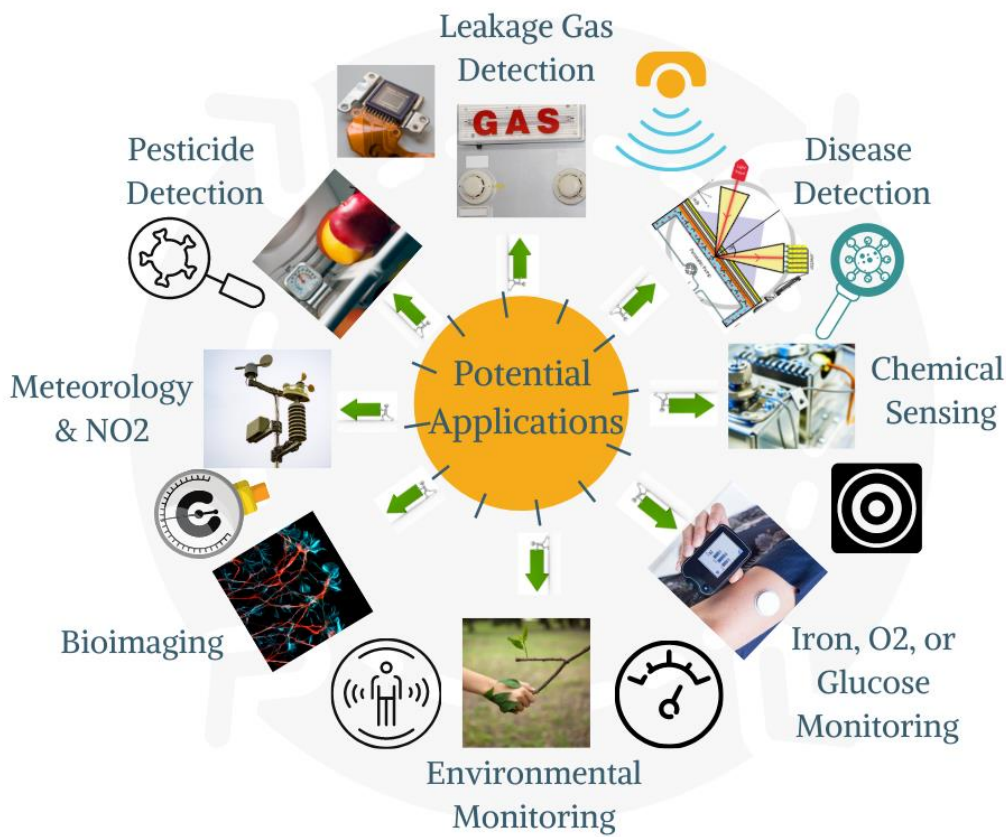

**Figure 9.** Potential applications of magneto-plasmonic sensors for the detection of (i) early disease (ii) chemicals, (iii) minerals, (iv) environmental monitoring, (v) bioimaging, (vi) meteorology and $NO_2$ (viii) pesticide and (viii) gas leakage.

Noteworthy, MO tuning combined with plasmonic behaviour, well-established manufacturing practices, and improved performance opens the door towards almost all types of the industrial sector. Figure 9 shows some of the target application areas of magneto-plasmonic sensors in the industry. Below we elaborate upon these areas.

1. **Early Disease Detection:** Many diseases caused by bacterial or viral infections can be prevented if detected early. Therefore, businesses need to develop unique healthcare devices targeted explicitly toward the diagnostic medicine sector. The market research also indicates a significant need for state-of-the-art healthcare technological devices dedicated to improving the quality of medical diagnosis within this region [182]. In this aspect, magneto-plasmonic technology can change the narrative in this industry in the areas such as detection, prevention/cure, medical diagnostics, and/or in the development of new diagnostic assays [183]. Moreover, healthcare providers, including doctors, hospitals, and clinics, can utilize magneto-plasmonic-based biosensors and AI & IoT-enabled smart devices [184] to detect diseases and/or collect real-time patient data for assessment. Furthermore, the unique properties of magneto-plasmonic structures can find application in bio-receptors antibodies, DNA, or more [185]. Users can use the magnetoplasmonic platform to study protein biomarkers due to their real-time and label-free detection such as cancer, cardiac diseases, etc. It is worth mentioning here that in 2021 a European company, BlueSense Diagnostics [186], exploiting MO effects as basic mechanisms in their sensors for early diagnosis of diseases, including COVID-19, was among the finalists at the European Patent Office Award [187], meaning that opto-magnetic functionalities can be really a game-changer in this sub-field of optical sensing and have a positive impact on societal challenges, such as pandemics prevention.

2. **Iron, $O_2$ or Glucose Monitoring:** SPR sensors have been routinely used for measuring blood metabolites such as lactate, glucose, creatinine, and urea, and iron or $O_2$ using both optical and electrochemical modes of transduction are being routinely used and developed in the point of care testing, laboratories, and self-testing for glucose monitoring. In the future, magnetoplasmonic sensors may replace the existing SPR sensors due to the superior sensitivity & detection limits in all cases.

3. **Chemical Sensing:** Chemical sensors are being used to measure and detect chemical characteristics in an analyte, e.g., breath analyzers, detection of $CO_2$, etc. The challenge is to address very small signals originating from a reaction between the chemical compounds. One can devise magnetoplasmonic configuration in different compositions and develop portable magnetoplasmonic sensors.

4. **Environmental Monitoring:** The SPR sensors have already been employed at various environmental conditions, including detecting the energetic material (also known as explosives) and monitoring RDX in the environment [188]. With the improved sensitivity and detection limit of the magnetoplasmonics sensors and the possibility of developing portable devices, magnetoplasmonic sensors may be employed to detect pathogenic species, heavy metals, and toxins in water with improved accuracy and without the need to access laboratory facilities, for example.

5. **Nanomedicine/Bioimaging**: The integration of magnetic and plasmonic features into a single nanostructure opens new opportunities in nanomedicine. Especially, magneto-plasmonic nanomaterials are excellent candidates as imaging probes for multimodal imaging that uses imaging modality to provide more precise images than a single modality, which helps to improve accuracy in disease diagnostics and provides more functional information [189].

6. **Meteorology and $NO_2$**: During harsh environmental conditions such as high humidity and high temperature, the threat of toxic gas exposure, such as $NO_2$ in the atmosphere, cannot be ignored. It can degrade air quality as well as cause severe damage such as inflammation to the human lung. Also, to detect $NO_2$ accurately and efficiently, one requires highly sensitive sensor configurations. In this aspect, magnetoplasmonic sensors can potentially replace the existing SPR sensors. With climate change being a huge issue, many users can also be aided to make more accurate predictions as these sensors can be optimized to detect changes in humidity or $NO_2$ in the atmosphere in harsh environmental conditions [190].

Some potential application sectors include government agencies (e.g., Environment Canada) and Meteorology companies worldwide.

7. **Agriculture/Pesticide**: Although there are no known links between COVID-19 or its new variants and the food or food packaging, scientists and food safety authorities worldwide have been monitoring potential food safety risks linked with COVID19 and cautioning the public to handle the safe processing of food products. Many other foodborne diseases are also emerging daily, and as a result, food safety is becoming essential worldwide more than ever [191]. In this context, companies and vendors must provide quality food products to minimize health risks. Although traditional methods such as plate count and polymerase chain reaction are accurate and robust, they are costly and time-consuming [192]. In this regard, the magneto-plasmonic sensors can be used to evaluate chemical contaminants/toxins, additives, and bioactive components in food and detect high and low-affinity small molecule and bacterial pathogens, the amount of methane present in beer and other food safety-related substances (e.g., bacterial pathogens in food samples in the field, pesticide detection, etc.). Likewise, they can be used as highly selective foodborne pathogens monitor due to their higher sensitivity and detection limits.

8. **Gas-leak Detection**: One area where SPR sensors may shine is detecting gases with extremely low concentrations [193]. Magneto-plasmonic sensors can be reliably employed to detect toxin gases at room temperature, for example.

Additional areas where magneto-plasmonic sensors may find applications include material characterization. For example, they can be used to monitor film thicknesses where the sensitivity response can change according to the thickness of the film. In addition, the sensors may find applications in automotive, manufacturing, aviation, marine, telecom, computer hardware, and space industries, to name a few. New configurations can be created where magnetic nanoparticles are suitably arranged across the structure's surface with which incident EM radiation interacts or the same can be achieved using an array of nanoholes imprinted across the interacting surface. Incorporating such enhancements to a substrate can offer the opportunity to detect, potentially, a single molecule. The high sensitivity of magneto-plasmonic sensors is of great attraction to sensing applications, particularly biomolecules.

**VI. Conclusions**

In this Perspective, we have showcased the state-of-the-art of magnetophotonic sensing in academic literature. Moreover, we have given a preliminary vision of this research field as a possible game-changer in many industrial areas where highly sensitive functionalities are needed for sensing and other types of applications, particularly magnetometry. We have tried to capture the most essential works in the field and provided our vision on advantages and limitations of using different magnetophotonic devices instead of conventional photonic or plasmonic structures. As the physical principles of magnetoplasmonics and magnetophotonics are more firmly established, smart approaches to increase the performance of magnetic field-based optical devices are required, both through transformative, out of the box thinking that can define new paradigms for materials and sensing schemes, as well as through more specific, application-oriented optimization. While magnetoplasmonic structures made of metallic materials show a higher sensitivity, for instance, for single-molecule detection, due to their very small detection volumes, the low-quality factor of the resonance might be a limiting factor. Non-metallic conductors, such as heavily doped semiconductors, could offer a solid alternative to metal-based magnetoplasmonics. On the other hand, hybrid metallic/dielectric systems might be a solution in the short run, but all-dielectric magnetophotonic structures might boost the research field in the long run. While still in its infancy, magnetophotonics has proven promising for sensing and magnetometry.

**Acknowledgements**

The authors thank Prof. Osvaldo N. Oliveira Jr, who helped us in the proof-reading of the manuscript. NM acknowledges support from the Luxembourg National Research Fund (Grant No. C19/MS/13624497 'ULTRON'), the European Commission under the FETOPEN-01-2018-2019-2020 call (Grant No. 964363 'ProID') and the FEDER Programme (Grant No. 2017-03-022-19 'Lux-Ultra-Fast'). J.R.M.-S. acknowledges financial support from the Brazilian agencies CNPq (429496/2018-4, 305958/2018-6) and RNP, with resources from MCTIC, Grant No. 01250.075413/2018-04, under the Radiocommunication Reference Center (Centro de Referência em Radiocomunicações - CRR) project of the National Institute of Telecommunications (Instituto Nacional de Telecomunicações - Inatel), Brazil. D.O.I. and V.I.B. acknowledge support from the Ministry of Science and Higher Education of the Russian Federation, Megagrant project N 075-15-2019-1934 and Russian Foundation of Basic Research project N 18-29-20113. We thank Zeynep Celik for the help with Fig. 9.

**Data Availability**

Data sharing is not applicable to this article as no new data were created or analyzed in this study.